# A Novel Approach to Fast Calculation of High-Order Q-Cumulants

L. Nađđerđ[1,2], J. Milošević[1,2], D. Devetak[1,2], F. Wang[4] and X. Zhu[3,2]
[1]*Department of physics, "VINČA" Institute of Nuclear Science - National Institute of the Republic of Serbia, University of Belgrade, Mike Petrovića Alasa 12-14, Vinča 11351, Belgrade, Serbia.*
[2]*Strong coupling Physics International Research Laboratory, Huzhou University, Huzhou, Zhejiang 313000, P. R. China.*
[3]*College of Science, Huzhou University, Huzhou, Zhejiang 313000, P. R. China.*
[4]*Department of Physics and Astronomy, Purdue University, Indiana 47907, USA.*

## Abstract

The method of Q-cumulants has been shown as a powerful tool to study the fine details of the azimuthal anisotropies in high-energy nucleus-nucleus collisions. A new method for the fast calculation of arbitrary order Q-cumulant $v_n\{2k\}$ values, based on the partition of a non-negative integer $l \leq m$ for calculation of the $2m$-particle azimuthal correlations is presented in this paper. Unlike the standard Q-cumulants method in which the calculation of high-order multi-particle calculations is impractical, the newly proposed method enables easy calculation. The validity of the method is proven via a toy model that uses the elliptic power distribution to simulate anisotropic emission of particles. The method enables the study of fine details of the $v_2$ distribution, such as higher-order central moments of the $v_2$ distribution, as well as the hydrodynamic behavior of the Quark-Gluon Plasma.

**Keywords:** Quark gluon plasma, Azimuthal anisotropies, Q-cumulants, Integer partition

## 1 Introduction

The Quark-Gluon-Plasma (QGP), a system consisting of strongly coupled quarks and gluons, is created in ultra-relativistic nuclear collisions [1-6]. The almond-like shape of the overlapping region in a nucleus-nucleus collision characterizes the initial spatial anisotropy. Due to this spatial anisotropy an anisotropic pressure gradient is built causing a collective azimuthally anisotropic expansion of the formed system. This phenomenon, called anisotropic collective flow, has been observed at the BNL AGS (E877) [7], the CERN SPS (NA49, WA98, CERES) [8-10] and the RHIC and LHC [6, 11-13]. The anisotropic collective flow is quantified by Fourier harmonics $v_n$ [14] and is dominated by second-order elliptic flow because of the almond-shape overlap geometry [15]. The fluctuating positions of the nucleons in the overlapping region cause the appearance of the higher-order Fourier harmonics $v_n$ ($n > 2$). The most pronounced higher-order Fourier harmonic is the triangular flow ($v_3$).

A few methods have been developed to measure magnitudes of the Fourier harmonics [15-17]. The main challenge is to suppress short-range correlations arising from jets and resonance decays, commonly referred to as non-flow effects, to reveal genuine collective flow of the QGP. The idea of using cumulants to perform measurements of azimuthal anisotropies was first proposed in [18, 19]. The method is based on cumulant expansion of multiparticle azimuthal correlations and uses the formalism of generating functions. The method mitigates contributions from non-flow effects by separating analyzed particles in pseudorapidity. A significant improvement of the method, called the Q-cumulants method [20], in principle, allows a fast calculation of all multi-particle correlations. Using this method, one can extract central moments of the $v_n$ distribution, for example skewness and kurtosis [21-23], to study hydrodynamic behavior of the QGP. The extraction of even higher-order central moments of the flow distribution requires higher-order Q-cumulants, which makes their measurement worthwhile. In practice, however, a further increase of the order of Q-cumulants, following the formalism given in Ref. [20], becomes difficult. Therefore, we developed a new approach for calculation of high-order Q-cumulants.

The rest of the paper is organized as follows. Section 2 describes basic variables of the Q-cumulant method. Section 3 presents foundation of the method. Section 4 demonstrates the validity of the method. A summary is given in section 5. A ROOT [24] code for calculating high-order Q-cumulants is given in the Appendix.

## 2 Basics of the Q-cumulants

The starting observable is flow vector Q defined as:

$$Q_n = \sum_{j=1}^{M} \mathrm{e}^{in\phi_j} \ , \tag{1}$$

where $n$ is the order of Fourier harmonics, $M$ is the multiplicity (the number of particles detected in an event), and $\phi_j$ is the azimuthal angle of the $j$-th particle in the laboratory coordinate system. Within the method, the $2m$-th power of the magnitude of the flow vector, $|Q_n|^{2m}$:

$$\left|Q_n\right|^{2m} = \sum_{j_1,\ldots,j_{2m}=1}^{M} \mathrm{e}^{in\left(\phi_{j_1}+\ldots+\phi_{j_m}-\phi_{j_{m+1}}-\ldots-\phi_{j_{2m}}\right)} \ , \tag{2}$$

is decomposed into off diagonal terms with $2m$, $2m$-1, … different indices up to the diagonal term with $2m$ equal indices. The first term of the decomposition, with $2m$ different indices, is proportional to the $2m$-particle azimuthal correlations denoted as $\langle 2m \rangle$:

$$\sum_{j_1\neq\ldots\neq j_{2m}=1}^{M} \mathrm{e}^{in\left(\phi_{j_1}+\ldots+\phi_{j_m}-\phi_{j_{m+1}}-\ldots-\phi_{j_{2m}}\right)} \equiv P_{M,2m} \cdot \langle 2m \rangle , \tag{3}$$

where $P_{M,2m}$ is the number of $2m$-particle distinct combinations in an event with multiplicity $M$:

$$P_{M,2m} = \frac{M!}{(M-2m)!} \, . \tag{4}$$

In the case of full decomposition, $|Q_n|^{2m}$ is expressed in readily calculable terms of powers of the flow vector given with Eq. (1) along with the anticipated $2m$-particle azimuthal

correlations introduced in Eq. (3). A thorough derivation with detailed examples is presented in Ref. [25]. The analytical decomposition of $|Q_n|^{2m}$ for $m > 4$, however, becomes tedious.

In this paper we present a numerical method of decomposition of $|Q_n|^{2m}$ that, with the use of modern computers, enables one to easily obtain analytical expressions of multi-particle azimuthal correlations of higher orders.

## 3 Foundation of the algorithm

In our previous work [26] we have shown that the 2*m*-particle azimuthal correlations $\langle 2m \rangle$ might be presented as a linear combination of basis vectors of a finite-dimensional vector space. In a comprehensive form this reads as:

$$P_{M,2m}\langle 2m\rangle = \sum_{l=0}^{m}\left[ f^{(m,l)}\sum_{i=1}^{\delta_l} N_i^{(m,l)}\mathbf{e}_i^{(l)}\right], \tag{5}$$

where $N_i^{(m,l)}$ are integer numbers, $\mathbf{e}_i^{(l)}$ are basis vectors of the corresponding $\delta_l$-dimensional subspace $W^{\delta_l}$. $f^{(m,l)}$ are integer functions of multiplicity $M$ given by

$$f^{(m,l)} = \begin{cases} (M-2l)\prod_{j=m+l+1}^{2m-1}(M-j), & \text{for} \quad l=\{0,\ldots,m-2\} \\ M-2l, & \text{for} \quad l=m-1 \\ 1, & \text{for} \quad l=m \end{cases} . \tag{6}$$

Basis vectors $\mathbf{e}_i^{(l)}$ consist of an appropriate products of flow vectors, $Q_{a_1n}Q_{a_2n}Q_{a_3n}\cdots$, with subscripts $a_1n$, $a_2n$, $a_3n,\ldots$, which correspond to the partition of a non-negative integer $l = a_1 + a_2 + a_3 + \cdots$. Here, we employed a convenient symbolic notation introduced in Refs. [20, 25]:

$$Q_{a\cdot n} \equiv \sum_{j=1}^{M} \mathrm{e}^{a\cdot in\phi_j}, a \in \{1, 2, 3, \ldots\}. \tag{7}$$

A partition of a non-negative integer $l$ consists of $v$ distinct parts $b_j$, each having the corresponding multiplicity $\mu_j$. For example, one of the partitions of number 12 is 3+3+2+1+1+1+1. It has $\nu = 3$ distinct parts $b_1 = 3$, $b_2 = 2$, and $b_3 = 1$, each having the corresponding multiplicities $\mu_1 = 2$, $\mu_2 = 1$, and $\mu_3 = 4$. In the multiplicity notation this might be written as $3^2 2^1 1^4$. Therefore, a composition of flow vectors consists of $v$ distinct sub-compositions, where each sub-composition is a product of $\mu_j$ number of flow vectors each having the subscript proportional to the same part $b_j$:

$$\underbrace{Q_{b_1n}Q_{b_1n}Q_{b_1n}\cdots}_{\mu_1}\underbrace{Q_{b_2n}Q_{b_2n}\cdots}_{\mu_2}\cdots\quad\cdots\underbrace{Q_{b_vn}Q_{b_vn}\cdots}_{\mu_v} = \prod_{j=1}^{v} Q_{b_jn}^{\mu_j}. \tag{8}$$

There are $p(l)$ distinct ways of representing a non-negative integer $l$ as a sum of non-negative integers (partition function of a non-negative integer $l$ = {0, 1, 2, 3, 4, 5 ...} is $p(l)$ = {1, 1, 2, 3, 5, 7 ...}). To obtain all basis vectors $\mathbf{e}_i^{(l)}$ one needs to combine each composition of the flow vectors by the complex conjugates of each composition which gives in total $p(l)\cdot p(l)$ super-compositions. Since a basis vector is represented only by the real value of a super-composition, the real value of its complex-conjugate represents the same basis vector, therefore one gets only $\delta_l = p(l)\left[p(l)+1\right]/2$ mutually independent basis vectors which span the whole subspace $W^{\delta_l}$. The total vector space of the 2*m*-particle correlations $V^{d_m}$ consists of $m+1$ subspaces, $V^{d_m} = \bigoplus_{l=0}^{m} W^{\delta_l}$, where $W^{\delta_l}$ is spanned by basis vectors which contain super-compositions of flow vectors that correspond to the partition of only the integer *l*. The dimension of the total vector space is the sum of the dimensions of all subspaces $d_m = \sum_{l=0}^{m} \delta_l$.

For example, vector space in the case of 6-particle azimuthal correlations ($m$ = 3), $V^{d_{m=3}}$, consists of four subspaces ($l$ = {0, 1, 2, 3} $\Rightarrow$ $p(l)$ = {1, 1, 2, 3} $\Rightarrow$ $\delta_l$ = {1, 1, 3, 6}) spanned by the following basis vectors $\mathbf{e}_i^{(l)}$ (Table 3 in Ref. [26]):

$$
\begin{aligned}
&W^{\delta_0} \Leftrightarrow \mathbf{e}_1^{(0)} \equiv 1 \\
&W^{\delta_1} \Leftrightarrow \mathbf{e}_1^{(1)} \equiv \mathrm{Re}(Q_n Q_n^*) \\
&W^{\delta_2} \Leftrightarrow \mathbf{e}_1^{(2)} \equiv \mathrm{Re}(Q_n Q_n Q_n^* Q_n^*),\, \mathbf{e}_2^{(2)} \equiv 2\,\mathrm{Re}(Q_{2n} Q_n^* Q_n^*),\, \mathbf{e}_3^{(2)} \equiv \mathrm{Re}(Q_{2n} Q_{2n}^*) \\
&W^{\delta_3} \Leftrightarrow \begin{cases} \mathbf{e}_1^{(3)} \equiv \mathrm{Re}(Q_n Q_n Q_n Q_n^* Q_n^* Q_n^*),\, \mathbf{e}_2^{(3)} \equiv 2\,\mathrm{Re}(Q_{2n} Q_n Q_n^* Q_n^* Q_n^*),\, \mathbf{e}_3^{(3)} \equiv \mathrm{Re}(Q_{2n} Q_n Q_n^* Q_{2n}^*) \\ \mathbf{e}_4^{(3)} \equiv 2\,\mathrm{Re}(Q_{3n} Q_n^* Q_n^* Q_n^*), \quad \mathbf{e}_5^{(3)} \equiv 2\,\mathrm{Re}(Q_{3n} Q_n^* Q_{2n}^*), \quad \mathbf{e}_6^{(3)} \equiv \mathrm{Re}(Q_{3n} Q_{3n}^*) \end{cases}
\end{aligned}
\quad , \quad (9)
$$

with total dimension of $d_3$ = 11. In Eq. (9), we doubled the real values of the asymmetric super-compositions (those with non-equal $Q$- and $Q^*$-parts of the super-compositions) in order to simplify the formula for determining the integers $N_i^{(m,l)}$.

In our previous work we obtained the integers $N_i^{(m,l)}$ by solving appropriate systems of linear algebraic equations [26]. By doing further investigations we revealed some interesting features of the Eq. (5) shown in Fig. 1 and Fig. 2, which guided us to obtain the explicit formula for determining the integers $N_i^{(m,l)}$ given by the Eq. (12) and Eq. (15). To illustrate this, we will write the first three well known multi-particle azimuthal correlations [20]:

$$
\begin{aligned}
P_{M,2}\langle 2\rangle &= |Q_n|^2 - M \\
P_{M,4}\langle 4\rangle &= |Q_n|^4 - 2\,\mathrm{Re}(Q_{2n} Q_n^* Q_n^*) + |Q_{2n}|^2 - 4(M-2)|Q_n|^2 + 2M(M-3) \\
P_{M,6}\langle 6\rangle &= |Q_n|^6 - 6\,\mathrm{Re}(Q_{2n} Q_n Q_n^* Q_n^* Q_n^*) + 9|Q_{2n}|^2|Q_n|^2 + 4\,\mathrm{Re}(Q_{3n} Q_n^* Q_n^* Q_n^*) - \\
&\quad -12\,\mathrm{Re}(Q_{3n} Q_n^* Q_{2n}^*) + 4|Q_{3n}|^2 - 9(M-4)|Q_n|^4 + 18(M-4)\,\mathrm{Re}(Q_{2n} Q_n^* Q_n^*) - \\
&\quad -9(M-4)|Q_{2n}|^2 + 18(M-2)(M-5)|Q_n|^2 - 6M(M-4)(M-5)
\end{aligned}
\quad . \quad (10)
$$

Terms which contain partitions of the highest integer in each of these multi-particle correlations might be written in comprehensive form as follows:

$$
\begin{aligned}
&\sum_{i=1}^{\delta_1} N_i^{(1,1)}\mathbf{e}_i^{(1)} \equiv |Q_n|^2 \\
&\sum_{i=1}^{\delta_2} N_i^{(2,2)}\mathbf{e}_i^{(2)} \equiv |Q_n|^4 - 2\,\mathrm{Re}(Q_{2n}Q_n^*Q_n^*) + |Q_{2n}|^2 \\
&\sum_{i=1}^{\delta_3} N_i^{(3,3)}\mathbf{e}_i^{(3)} \equiv \begin{cases} |Q_n|^6 - 3\cdot 2\,\mathrm{Re}(Q_{2n}Q_nQ_n^*Q_n^*Q_n^*) + 9|Q_{2n}|^2|Q_n|^2 + 2\cdot 2\,\mathrm{Re}(Q_{3n}Q_n^*Q_n^*Q_n^*) \\ \qquad - 6\cdot 2\,\mathrm{Re}(Q_{3n}Q_n^*Q_{2n}^*) + 4|Q_{3n}|^2 \end{cases} \\
&\qquad \vdots \quad \equiv \quad \vdots
\end{aligned} \quad . \tag{11}
$$

The integer coefficients $N_i^{(l,l)}$ are deducible from their known values for $l = 0,\ldots,7$, $i = 1,\ldots,\delta_l$, [20, 26], and are given by the following arithmetic function of the integer partition parameters *b*, μ, and *v*, of the basis vectors:

$$
N_i^{(l,l)} = \frac{(-1)^{\sum_{j=1}^{v_i}\mu_{i,j}}\, l!}{\prod_{j=1}^{v_i}\left(b_{i,j}{}^{\mu_{i,j}}\cdot \mu_{i,j}!\right)} \frac{(-1)^{\sum_{j=1}^{\bar{v}_i}\bar{\mu}_{i,j}}\, l!}{\prod_{j=1}^{\bar{v}_i}\left(\bar{b}_{i,j}{}^{\bar{\mu}_{i,j}}\cdot \bar{\mu}_{i,j}!\right)}, \tag{12}
$$

where $v_i$ counts the number of distinct parts of the integer partition in the Q-part of the *i*-th super-composition (*i*-th basis vector), $b_{i,j}$ is the *j*-th distinct part of the partition, and $\mu_{i,j}$ is its multiplicity. $\bar{v}_i$, $\bar{b}_{i,j}$, and $\bar{\mu}_{i,j}$ are corresponding numbers in the $Q^*$-part of the same super-composition *i*.

For example, the integer coefficient $N_2^{(3,3)}$ of the basis vector $\mathbf{e}_2^{(3)} \equiv 2\,\mathrm{Re}(Q_{2n}Q_nQ_n^*Q_n^*Q_n^*)$ is:

$$
N_2^{(3,3)} = \frac{(-1)^{1+1}3!}{(2^1\cdot 1!)(1^1\cdot 1!)}\frac{(-1)^3 3!}{(1^3\cdot 3!)} = -3. \tag{13}
$$

Another example is the integer coefficient $N_5^{(3,3)}$ of the basis vector $\mathbf{e}_5^{(3)} \equiv 2\,\mathrm{Re}(Q_{3n}Q_n^*Q_{2n}^*)$:

$$
N_5^{(3,3)} = \frac{(-1)^1 3!}{(3^1\cdot 1!)}\frac{(-1)^{1+1}3!}{(1^1\cdot 1!)(2^1\cdot 1!)} = -6. \tag{14}
$$

Each basis vector $\mathbf{e}_i^{(l)}$ is at the same time a basis vector for all multi-particle correlations whose order is higher than *l* [26]. However, in each multi-particle correlation it enters with a different integer coefficient $N_i^{(m,l)}$. For example in (Eq. 10), the basis vector $|Q_n|^2$ has the following integer coefficients $1, -4,$ and 18, respectively. In order to reveal the integer coefficients $N_i^{(m,l)}$, we divided each observable $\sum_{i=1}^{\delta_l} N_i^{(m,l)}\mathbf{e}_i^{(l)}/m!$ by $\sum_{i=1}^{\delta_l} N_i^{(l,l)}\mathbf{e}_i^{(l)}/l!$ and obtained the following results shown in Fig. 1.

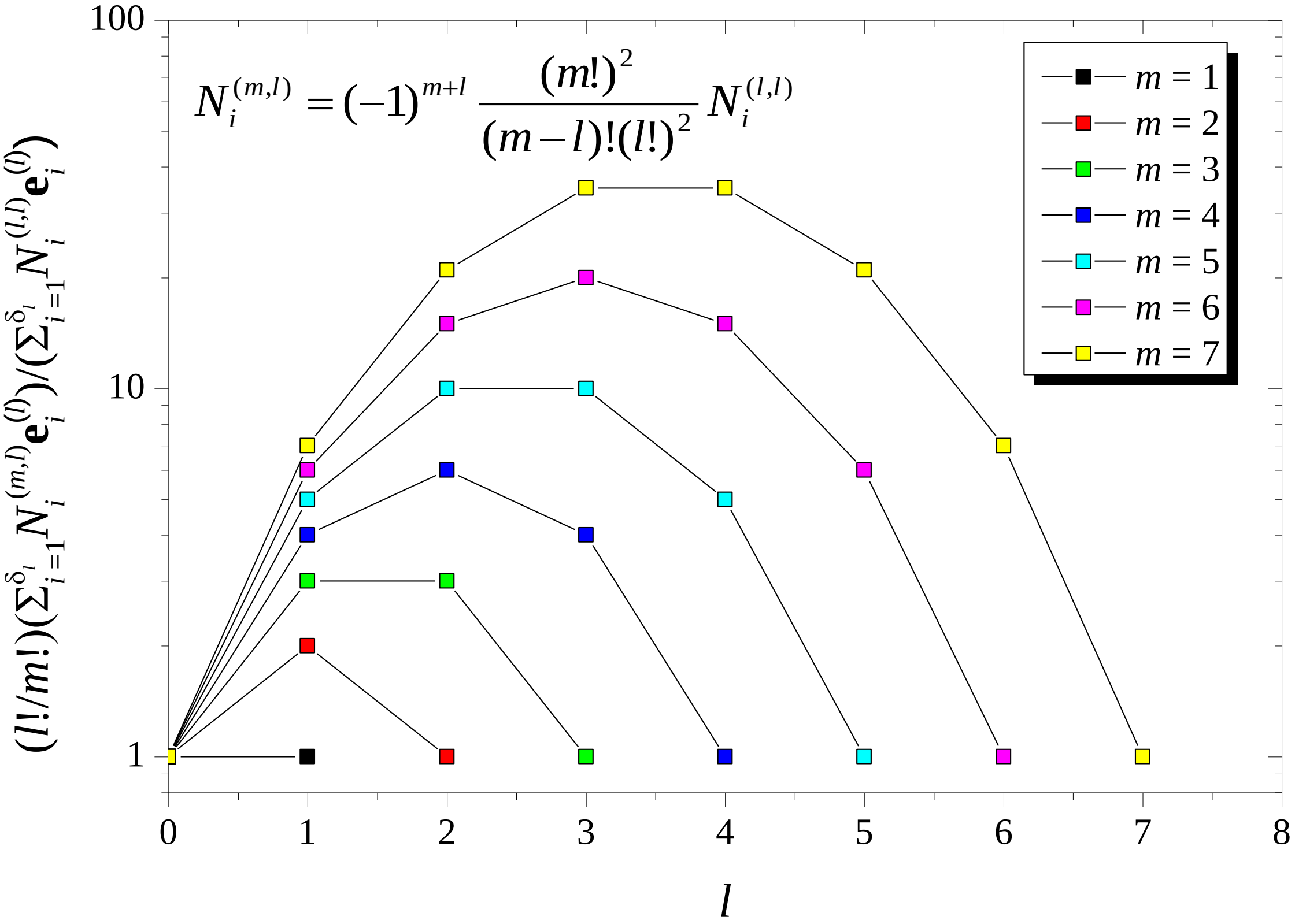

Fig.1 The values of the observables $(l!\sum_{i=1}^{\delta_l} N_i^{(m,l)}\mathbf{e}_i^{(l)})/(m!\sum_{i=1}^{\delta_l} N_i^{(l,l)}\mathbf{e}_i^{(l)})$, $l$ = 0, 1, 2,…,7 for $m$ = 1, 2, 3,…,7. The straight lines are to guide the eye.

The described features helped us to reveal the following binomial-like (Fig. 1) dependence of the $N_i^{(m,l)}$ integer coefficients on $m$ and $l$:

$$N_i^{(m,l)} = (-1)^{m+l}\frac{(m!)^2}{(m-l)!(l!)^2}N_i^{(l,l)}, \quad (15)$$

In practice, one is not always interested in intermediate results of calculation such as the terms written in Eq. (11), but rather the final results. In this case, in the course of building up the code for calculation of Eq. (5) it might be reduced by the same factor $(l!)^2$, which is present in both equations, Eq.(15) and Eq. (12).

In the case when the multiplicity of an event is equal to the Q-cumulant ordering number, $M = m$, the part of the Eq. (5) for which $l = m$, is always equal to $(M!)^2$:

$$\sum_{l=m}^{m}(f^{(m,l)}\sum_{i=1}^{\delta_l} N_i^{(m,l)}\mathbf{e}_i^{(l)})\Big|_{M=m} = \sum_{i=1}^{\delta_m} N_i^{(m,m)}\mathbf{e}_i^{(m)}\Big|_{M=m} = (M!)^2. \quad (16)$$

This finding, which remained unnoticed so far, is presented in Fig. 2.

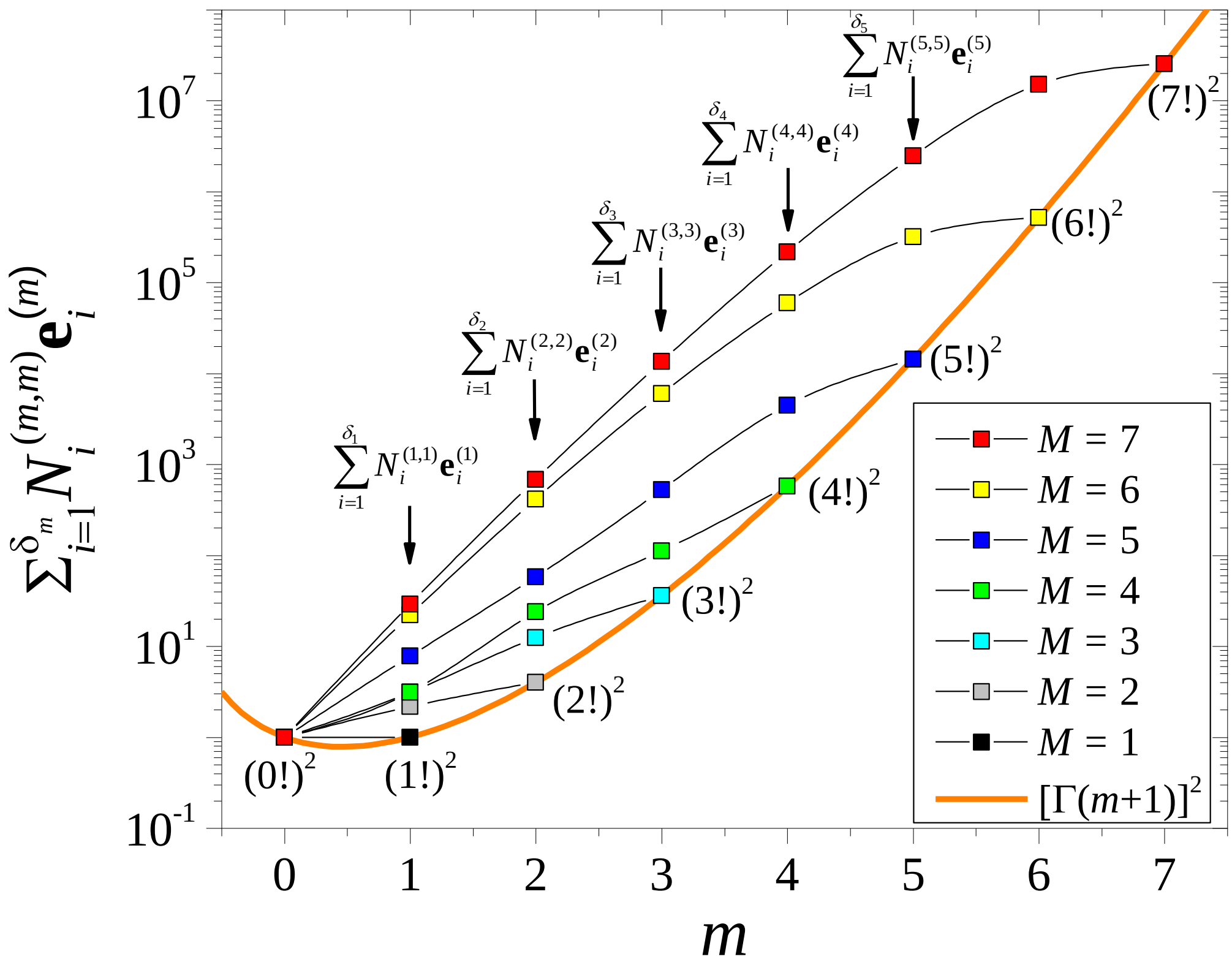

Fig.2 The values of the observables $\sum_{i=1}^{\delta_m} N_i^{(m,m)} \mathbf{e}_i^{(m)}$, $m = 1,\ldots,7$, for multiplicities from $M = 1$ to $M = 7$. The spline lines are to guide the eye.

Another interesting feature is obtainable by dividing each observable $\sum_{i=1}^{\delta_m} N_i^{(m,m)} \mathbf{e}_i^{(m)}$ by the corresponding $(m!)^2$. The obtained results are shown in Fig. 3 for $0 \le m \le 7$, in addition to the results for $8 \le m \le 20$ after applying the equations Eq. (12) and Eq. (15), which gives further confirmation for the correctness of these two formulas.

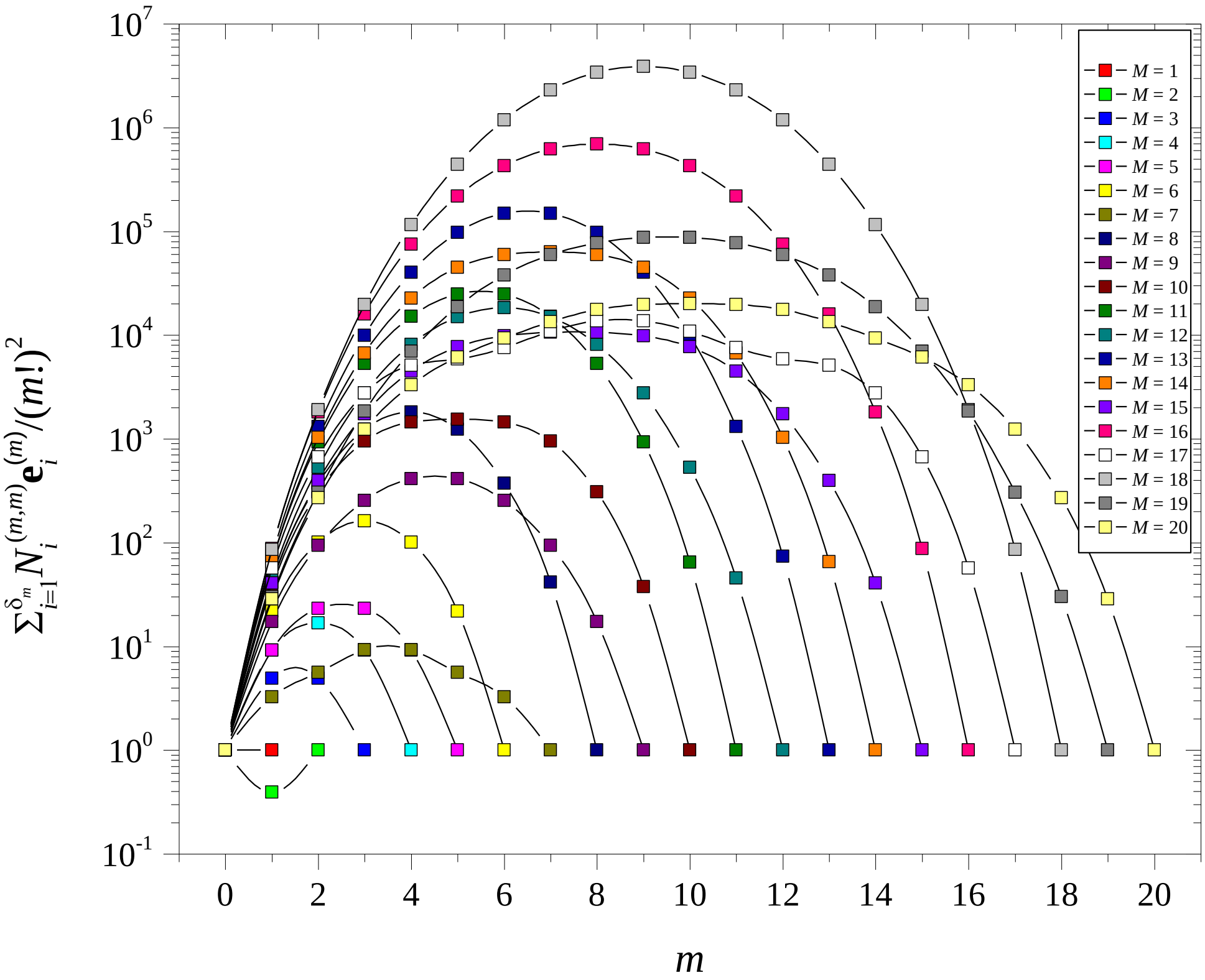

Fig.3 The values of the observables $\sum_{i=1}^{\delta_m} N_i^{(m,m)}\mathbf{e}_i^{(m)}/(m!)^2$, $m$ = 1, 2, 3,…, 20, for multiplicities $M$ = 1, 2, 3,…, 20. The spline lines are to guide the eye.

Finally, having all the necessary components for calculation of the 2$m$-particle azimuthal correlations $\langle 2m \rangle$ given by Eq.(5), one can compute the weighted average over all events $\langle\langle 2m \rangle\rangle$ (given by Eq. (1) in Ref. [27]). Then, one can use the recurrence relation to calculate the Q-cumulants $c_n\{2k\}$ of any order by knowing all the Q-cumulants of lower orders [27]:

$$c_n\{2k\} = \langle\langle 2k \rangle\rangle - \sum_{m=1}^{k-1} \binom{k}{m}\binom{k-1}{m} \langle\langle 2m \rangle\rangle c_n\{2k-2m\}. \quad (17)$$

The Q-cumulant based flow harmonics $v_n\{2k\}$ ($k$ = 1, 2, 3,…) are calculated using the following equation [28]:

$$v_n\{2k\} = \sqrt[2k]{a_{2k}^{-1} c_n\{2k\}}, \quad (18)$$

where coefficients $a_{2k}$ are obtainable by recursion relation [28]:

$$a_{2k} = 1 - \sum_{m=1}^{k-1} \binom{k}{m}\binom{k-1}{m} a_{2k-2m}\text{, with: } a_2 = 1, \quad (19)$$

which enables easy calculation of high-order $v_n\{2k\}$.

It has been shown in Ref. [27] that the statistical uncertainties of the Q-cumulants, $c_n\{2k\}$, increase slightly, nearly negligible, with the cumulant order due to high correlations between the weighted averages of the multi-particle azimuthally correlations, $\langle\langle 2k \rangle\rangle$. In addition, the Q-cumulants of different orders are mutually highly correlated [23, 27], which makes their ratios to be potentially high-precision quantities, as will be shown in the next section.

## 4 Validation of the method using a toy model

In order to validate the above described method, the obtained expressions for the Q-cumulants up to the 40-th order are calculated using azimuthal angles simulated with a toy model. The initial eccentricity $\varepsilon_2$ distribution [29] is simulated using the elliptic power distribution with sets of parameters of different values, depend on the centrality, obtained by Glauber model for 5.02 GeV PbPb collisions [30]. Also, we varied the multiplicity of the events by Gaussian distribution in a range which corresponds to the centrality of a PbPb collision (shown in Table 1.). The scaling factor $\kappa_2$ between the elliptic flow and the initial eccentricity, $v_2 = \kappa_2 \varepsilon_2$, is chosen to imitate the centrality dependence of the elliptic flow $v_2$ measured in Ref. [23]. For each event of a given $v_2$, a simple distribution $1 + 2v_2 \cos(2\phi)$ is used to generate particle azimuthal angle $\phi$. For each centrality, $10^6$ events are simulated. The statistical uncertainties are estimated by bootstrapping method which has been proven in Ref. [27] to be in accordance with the analytical method where the proper weighting of the covariance matrix has been applied.

**Table 1** Set of parameters used in the toy-model simulations for each centrality class. Power, $\alpha$, and ellipticity, $\boldsymbol{\varepsilon_0}$, are Elliptic-power distribution parameters [29]. Mean multiplicity, $\mu$, and standard deviation of the multiplicity, $\sigma$, are Gaussian distribution parameters. Scaling factor, $\kappa_2$, is a linear response between elliptic flow and initial eccentricity. The parameters are chosen to emulate the real charged-particle azimuthal distributions at 5.02 TeV PbPb collisions.

| Bin # | Centrality class [%] | Power parameter $\alpha$ | Ellipticity parameter $\varepsilon_0$ | Mean multiplicity $\mu$ | Standard deviation $\sigma$ | Scaling factor $\kappa_2$ |
|---|---|---|---|---|---|---|
| 1 | 10-15 | 48.41 | 0.169 | 1254 | 95.62 | 0.3605 |
| 2 | 15-20 | 40.62 | 0.216 | 1058 | 85.11 | 0.3551 |
| 3 | 20-25 | 32.82 | 0.261 | 882.1 | 75.44 | 0.3427 |
| 4 | 25-30 | 26.26 | 0.296 | 728.0 | 66.76 | 0.3342 |
| 5 | 30-35 | 22.15 | 0.329 | 589.4 | 59.14 | 0.3227 |
| 6 | 35-40 | 17.23 | 0.349 | 468.1 | 51.43 | 0.3144 |
| 7 | 40-45 | 13.95 | 0.367 | 364.0 | 44.12 | 0.3030 |
| 8 | 45-50 | 11.49 | 0.385 | 276.5 | 37.28 | 0.2848 |
| 9 | 50-55 | 9.03 | 0.400 | 204.6 | 30.99 | 0.2640 |
| 10 | 55-60 | 6.97 | 0.402 | 147.2 | 25.30 | 0.2436 |

As the fluctuations in the initial state are not Gaussian, the $v_2\{2k\}$ values for $k > 1$ will not be the same. This will produce a splitting between different $v_2\{2k\}$ and they will be ordered as $v_2\{2k\} > v_2\{2(k + 1)\}$ for any $k > 1$. To make the splitting between the cumulants of different orders explicitly visible, we show in Fig. 4 the relative differences $(v_2\{2k\} - v_2\{40\})/v_2\{40\}$ ($k = 1,\dots,19$) as a function of centrality. The magnitudes in Fig. 4 presented with colored markers correspond to different cumulant orders calculated from the simulated data. On the other hand, the corresponding input values, obtained directly by applying the elliptic power distributions, are represented by small black squares connected by straight lines. In Fig. 4 the ordering is clearly seen, as well as the fine splitting between the cumulants of different orders. The relative difference between the cumulants decreases by about one order of magnitude for each increment of the order $k$. An excellent agreement between the black squares and the symbols confirms the correctness of the obtained expressions for the $2m$-particle azimuthal correlations $\langle 2m \rangle$. The statistical uncertainties are smaller than the symbol size. In the most central and the most peripheral collisions for the highest cumulant orders, there is a small difference between theoretical values and calculations from the simulated data. However, this difference is within the standard statistical uncertainty limits.

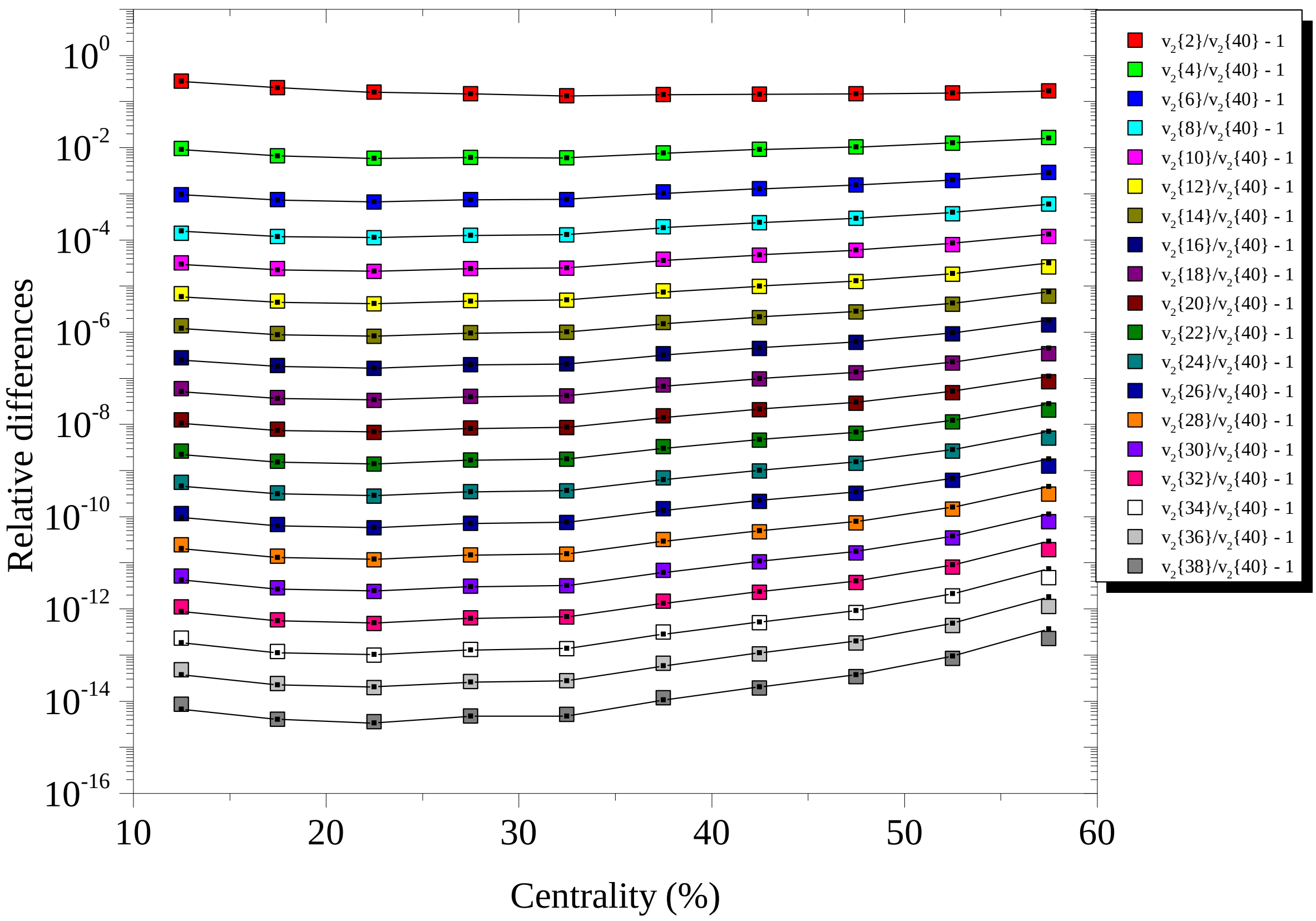

Fig.4 The relative differences $(v_2\{2k\} - v_2\{40\})/v_2\{40\}$ ($k = 1, 2, \dots, 19$) as a function of centrality are presented as colored squares. The input values, obtained by directly applying the elliptic power distributions, are represented by small black squares connected by straight lines. Statistical uncertainties are smaller than the symbol size.

**Conclusions**

A new method for the fast calculation of any order Q-cumulant $v_n\{2k\}$ values is presented in this paper. The method is based on the partition of a non-negative integer $l \leq m$ for calculation of the $2m$-particle azimuthal correlations. The standard Q-cumulants method, in principle, allows calculation of any order multi-particle calculations, but in practice it becomes very difficult to determine the analytical expressions of the cumulants that include azimuthal correlations of six or more particles. The validity of the method has been proven via a toy model that uses the elliptic power distribution to simulate anisotropic emission of particles. To study the fine details of the $v_2$ distribution, including higher-order central moments, as well as the hydrodynamic behavior of the QGP one needs higher-order Q-cumulant $v_n\{2k\}$ values. The proposed method enables their calculation.

## Acknowledgments

The authors acknowledge the support from Ministry of Science, Technological Development and Innovation of Republic of Serbia throughout the theme 0102502, Collaboration Projects JINR-Serbia_P23 and JINR-Serbia_P39, National Natural Science Foundation of China (Grant No. 12035006, 12075085, 12147219) and the U.S. Department of Energy (Grant No. de-sc0012910).

## References

[1] I. Arsene, et al., BRAHMS, Quark–gluon plasma and color glass condensate at RHIC? The perspective from the BRAHMS experiment, *Nucl. Phys. A* **757** (2005) 1, https://doi.org/10.1016/j.nuclphysa.2005.02.130, arXiv:nucl-ex/0410020.
[2] K. Adcox, et al., PHENIX, Formation of dense partonic matter in relativistic nucleus–nucleus collisions at RHIC: experimental evaluation by the PHENIX Collaboration, *Nucl. Phys. A* **757** (2005) 184, https://doi.org/10.1016/j.nuclphysa.2005.03.086, arXiv:nucl-ex/0410003.
[3] B. B. Back, et al., PHOBOS, The PHOBOS perspective on discoveries at RHIC, *Nucl. Phys. A* **757** (2005) 28, https://doi.org/10.1016/j.nuclphysa.2005.03.084, arXiv:nucl-ex/0410022.
[4] J. Adams, et al., STAR, Experimental and theoretical challenges in the search for the quark–gluon plasma: the STAR Collaboration's critical assessment of the evidence from RHIC collisions, *Nucl. Phys. A* **757** (2005) 102, https://doi.org/10.1016/j.nuclphysa.2005.03.085, arXiv:nucl-ex/0501009.
[5] G. Roland, K. Šafařík, and P. Steinberg, Heavy-ion collisions at the LHC, *Prog. Part. Nucl. Phys.* **77** (2014) 70, https://doi.org/10.1016/j.ppnp.2014.05.001
[6] B. Muller, J. Schukraft, and B. Wysłouch, First Results from Pb+Pb collisions at the LHC, *Ann. Rev. Nucl. Part. Sci.* **62** (2012) 361. https://doi.org/10.1146/annurev-nucl-102711-094910, arXiv:1202.3233 [hep-ex]
[7] E877 Collaboration (J. Barrette et al.), *Phys. Rev. Lett.* **73** (1994) 2532.
[8] NA49 Collaboration (T. Wienold), *Nucl. Phys. A* **610** (1996) 76.

[9] WA98 Collaboration (M. M. Aggarwal et al.), *Nucl. Phys. A* **638** (1998) 459.
[10] CERES Collaboration (F. Ceretto et al.), *Nucl. Phys. A* **638** (1998) 467.
[11] ATLAS Collaboration, Observation of a centrality-dependent dijet asymmetry in lead–lead collisions at $\sqrt{s_{NN}}$ = 2.76 TeV with the ATLAS detector at the LHC, *Phys. Rev. Lett.* **105** (2010) 252303,
https://doi.org/10.1103/PhysRevLett.105.252303, arXiv:1011.6182.
[12] ALICE Collaboration, Suppression of charged particle production at large transverse momentum in central Pb-Pb collisions at $\sqrt{s_{NN}}$ =2.76 TeV, *Phys. Lett. B* **696** (2011) 30,
https://doi.org/10.1016/j.physletb.2010.12.020, arXiv:1012.1004.
[13] CMS Collaboration, Observation and studies of jet quenching in PbPb collisions at $\sqrt{s_{NN}}$ = 2.76 TeV, *Phys. Rev. C* **84** (2011) 024906,
https://doi.org/10.1103/ PhysRevC.84.024906, arXiv:1102.1957.
[14] S. Voloshin and Y. Zhang, Flow study in relativistic nuclear collisions by Fourier expansion of azimuthal particle distributions, *Z. Phys. C* **70** (1996) 665,
https://doi.org/10.1007/s002880050141, arXiv: hep-ph/9407282.
[15] J.-Y. Ollitrault, Anisotropy as a signature of transverse collective flow, *Phys. Rev. D* **46** (1992) 229.
[16] J.-Y. Ollitrault, Determination of the reaction plane in ultrarelativistic nuclear collisions, *Phys. Rev. D* **48** (1993) 1132,
https://doi.org/10.1103/PhysRevD.48.1132, arXiv:hep-ph/9303247.
[17] A. M. Poskanzer and S. A. Voloshin, Methods for analyzing anisotropic flow in relativistic nuclear collisions, *Phys. Rev. C* **58** (1998) 1671,
https://doi.org/10.1103/PhysRevC.58.1671, arXiv:nucl-ex/9805001.
[18] N. Borghini, P. M. Dinh, and J.-Y.Ollitrault, A New method for measuring azimuthal distributions in nucleus-nucleus collisions, *Phys. Rev. C* **63** (2001) 054906,
https://doi.org/10.1103/PhysRevC.63.054906, arXiv:nucl-th/0007063.
[19] N. Borghini, P. M. Dinh, and J.-Y.Ollitrault, Flow analysis from multiparticle azimuthal correlations, *Phys. Rev. C* **64** (2001) 054901,
https://doi.org/10.1103/PhysRevC.64.054901, arXiv:nucl-th/0105040.
[20] A. Bilandzic, R. Snellings, and S. Voloshin, Flow analysis with cumulants: Direct calculations, *Phys. Rev. C* **83** (2011) 044913,
https://doi.org/10.1103/PhysRevC.83.044913, arXiv:nucl-ex/1010.0233.
[21] G. Giacalone, L. Yan, J. Noronha-Hostler, and J.-Y.Ollitrault, Skewness of elliptic flow fluctuations, *Phys. Rev. C* **95** (2017) 014913,
https://doi.org/10.1103/PhysRevC.95.014913, arXiv:nucl-th/1608.01823.
[22] R. S. Bhalerao, G. Giacalone, and J.-Y.Ollitrault, Kurtosis of elliptic flow fluctuations, *Phys. Rev. C***99** (2019) 014907,
https://doi.org/10.1103/PhysRevC.99.014907, arXiv.org/abs/1811.00837.
[23] CMS Collaboration, Higher-order moments of the elliptic flow distribution in PbPb collisions at $\sqrt{s_{NN}}$ = 5.02 TeV, *J. High Energy Phys.* **02** (2024) 106,
arXiv:2311.11370 [nucl-ex].
[24] R. Brun and F. Rademakers, ROOT - An Object Oriented Data Analysis Framework, Proceedings AIHENP'96 Workshop, Lausanne, Sep. 1996, *Nucl. Inst. & Meth. in Phys. Res. A* **389** (1997) 81-86.
[25] A. Bilandzic, Anisotropic flow measurements in ALICE at the large hadron collider, PhD thesis, Utrecht U., CERN-THESIS-2012-018, 2012.
[26] L. Nađđerđ, J. Milošević, D. Devetak and Fu-Qiang Wang, Decomposition of multi-particle azimuthal correlations in *Q*-cumulant analysis, *Chinese Phys. C* **47** (2023) 104107.

[27] L. Nadderd, J. Milosevic, and F. Wang, Statistical uncertainties of the $v_n\{2k\}$ harmonics from Q cumulants, *Phys. Rev. C* **104** (2021) 034906.
[28] CMS Collaboration, Probing hydrodynamics and the moments of the elliptic flow distribution in $\sqrt{s_{NN}}$ = 5.02 TeV lead-lead collisions using higher-order cumulants, CMS-PAS-HIN-21-010, (2022).
[29] L. Yan, J.-Y.Ollitrault, and A. M. Poskanzer, Eccentricity distributions in nucleus-nucleus collisions, *Phys. Rev. C* **90** (2014) 024903.
[30] N. Abbasi, D. Allahbakhshi, A. Davody, and S. F. Taghavi, Standardized cumulants of flow harmonic fluctuations, *Phys. Rev. C* **98** (2018) 024906.
[31] J. Kelleher and B. O'Sullivan, Generating All Partitions: A Comparison of Two Encodings, arXiv:0909.2331v2 [cs.DS] 2 May 2014.

## Appendix

```
void Qcumulants(){

Double_t mm = 20; // highest Q-cumulant order of the calculation (mm=1 is cumulant of the 2nd order)
Double_t E = 1000000; // number of events
Double_t Bootstrapps = 20; // number of bootstrappings (for uncertainty calculation)
// lower number of events requires higher number of bootstrappings for good estimations of the
uncertainties

Double_t alpha = 48.41, epsilon0 = 0.169, kappa2 = 0.3605;
// parameters of the elliptic-power distribution and linear response kappa2 (only for simulations)

Double_t mu = 1254, sigma = 95.62;
// parameters of the Gauss distribution for the multiplicity of an event (only for simulations)

vector <vector<Double_t>> XN(E, vector<Double_t>(mm+1)), XD(E, vector<Double_t>(mm+1));
vector <vector<Double_t>> YN(E, vector<Double_t>(mm+1)), YD(E, vector<Double_t>(mm+1));
// numerators and denominators for bootstrappings

vector <vector<Double_t>> Ave(Bootstrapps+1, vector<Double_t>(mm+1));
vector <vector<Double_t>> Cum(Bootstrapps+1, vector<Double_t>(mm+1));
vector <vector<Double_t>> sACum(Bootstrapps+1, vector<Double_t>(mm+1));
vector <vector<Double_t>> v2(Bootstrapps+1, vector<Double_t>(mm+1));
vector <vector <Double_t>> f(mm+1, vector <Double_t>(mm+1));
// function of multiplicity, f, given by Eq. (6)

vector <vector <Double_t>> Sumprodv2(mm+1, vector<Double_t>(mm+1)), Cov(mm+1, vector<Double_t>(mm+1));
vector <vector <Double_t>> b(0, vector <Double_t>(0)); // distinct parts of the partition
vector <vector <Double_t>> mult(0, vector <Double_t>(0)); // multiplicity of a partition
vector <vector<complex<Double_t>>> Q(mm+1, vector<complex<Double_t>>(6));
// first index counts the indices of the Q vector (Q_1, Q_2, Q_3,...), the second index counts
harmonics k0=1,2,3,4,5 (we avoid k0=0 but it is included in the dimensions of the vector = 6)

vector <Double_t> Num(mm+1), Den(mm+1);
// numerators and denominators of the 2m-particle azimuthal correlations <2m> for an event

vector <Double_t> sR(mm+1), SumYN(mm+1), SumYD(mm+1);
vector <Double_t> aa(mm+1), saa(mm+1), Sumv2(mm+1);
vector <complex<Double_t>> z(0), w(0);

TF1 *f1 = new TF1("f1","1.0+2.0*[1]*cos(1.0*x)+2.0*[2]*cos(2.0*x)", -TMath::Pi(), TMath::Pi());
// Fourier function (only for simulations)

TF1 *f2 = new TF1("f2","2*[0]*pow(1-pow([1],2),[0]+0.5)*x*pow(1-x*x,[0]-
1)/pow(1+[1]*x,2*[0]+1)*ROOT::Math::hyperg(0.5,2*[0]+1,1,(2*[1]*x)/(1+[1]*x))", 0, 1);
// elliptic-power distribution (only for simulations)

f2->SetParameter (0, alpha), f2->SetParameter (1, epsilon0);

for (Int_t ii=0; ii<E; ii++){
```

```
        if (ii%1000==0) cout <<ii<<endl;
        Double_t M = floor(gRandom->Gaus(mu, sigma)); // multiplicity of an event
        Double_t epsilonx = f2->GetRandom(); // eccentricity selected from elliptic-power distribution
(only for simulations)
        Double_t v_11 = 0, v_22 = kappa2*epsilonx; // elliptic flow takes values (directed flow set to
zero)

        f1->SetParameter(1,v_11), f1->SetParameter(2,v_22);

        for (Int_t m=1; m<=mm; m++) Q[m][2]=0.0; // Q vectors annulations, here we calculate only
elliptic flow (second index set to 2)

        for (Int_t jj=0; jj<M; jj++){
                Double_t phi = f1->GetRandom(); // azimuthal angles selected from Fourier function
(only for simulations)
                for (Int_t n0=1; n0<=mm; n0++){ // over Q-vector indices
                        for (Int_t k0=2; k0<=2; k0++){ // over Fourier harmonics, here we calculate
only elliptic flow (second index set to 2)
                        Q[n0][k0] += exp(complex<Double_t>(0.0, n0*k0*phi)); // Q vectors
                        }
                }
        }

        for (Int_t m=1; m<=mm; m++) Num[m]=0.0, Den[m]=0.0;

        for (Int_t m=1; m<=mm; m++){
                Den[m]=1;
                for (Int_t j=0; j<2*m; j++) Den[m] *= M-j;
// denominators of the 2m-particle azimuthal correlations <2m> for an event
                for (Int_t l=0; l<=m; l++){
                        if (l==0){
                        sR[l]=1;
                        }
                        else {
/*===================*/ Int_t j=0, k=1;
/*                   */ Double_t a[l];
/* integer partition */ a[0]=0, a[1]=l;
/*                   */ sR[l] = 0;
/* part of the code  */         while (k!=0){
/*                   */                 Int_t x = a[k-1] + 1, y = a[k] - 1;
/*      given in     */                 k -= 1;
/*                   */                 while (x<=y){
/*      Ref. [31]    */                         a[k] = x, y = y - x, k += 1;
/*                   */                 }
/*===================*/                 a[k] = x + y;
                                        z.resize(j+1);
                                        z[j]=1;
                                        Int_t v=0;
                                        b.resize(j+1, vector <Double_t> (mm+1));
                                        b[j][0] = a[0];
                                        mult.resize(j+1, vector <Double_t> (mm+1));
                                        mult[j][0]=0;
                                        for (Int_t i=0; i<=k; i++){
                                                if (b[j][v] < a[i]){
                                                        v += 1;
                                                        b.resize(j+1, vector <Double_t> (v+1));
                                                        b[j][v] = a[i];
                                                        mult.resize(j+1, vector <Double_t> (v+1));
                                                        mult[j][v] = 1;
                                                }
                                                else mult[j][v] += 1;

                                                z[j] *= Q[a[i]][2];
                                        }
                                        Double_t smult=0;
                                        Double_t pb=1;
                                        for (Int_t i=0; i<=v; i++){
                                        smult += mult[j][i];
                                        pb *= pow(b[j][i],mult[j][i])*TMath::Factorial(mult[j][i]);
                                        }
                                        w.resize(j+1);
                                        w[j] = z[j]*pow(-1,smult)/pb;
                                        for (Int_t s=0; s<=j; s++){
                                                if (s==j) sR[l] += pow(abs(w[s]),2);
```

```
                                                else sR[l] += 2*real(w[j]*conj(w[s]));
                                        }
                                        j += 1;
                                }
                        }

                        if (l==m) f[m][l]=1;
                        else if (l==m-1) f[m][l] = M-2*l;
                        else {
                                f[m][l] = (M-2*l);
                                for (Int_t j=m+l+1; j<2*m; j++) f[m][l] *= (M-j);
                        }
                Num[m] += pow(-1,m+l)*pow(TMath::Factorial(m),2)/TMath::Factorial(m-l)*f[m][l]*sR[l];
                }
                XN[ii][m] = Num[m], XD[ii][m] = Den[m];
        }
}

cout << endl;

Int_t p = 2, zz = 2;
aa[1] = 1;

for (Int_t i=1; i<mm; i++){
        for (Int_t j=1; j<zz; j++){
        saa[zz] += TMath::Binomial(zz,j)*TMath::Binomial(zz-1,j)*aa[zz-j];
        }
        aa[p] = 1 - saa[zz]; // coefficients given by Eq. (19)
        p += 1;
        zz = p;
}

cout << " wait for bootstrapping ... " << endl;

for (Int_t l=0; l<=Bootstrapps; l++){
        if (l%100==0) cout <<l<<endl;
        for (Int_t m=1; m<=mm; m++) SumYN[m]=0.0, SumYD[m]=0.0;

        for (Int_t nn=0; nn<E; nn++){
                Int_t n = nn * (1 + TMath::Sign(1, -l))/2 + (1 - TMath::Sign(1, -l))/2 * gRandom-
>Integer(E);
                for (Int_t m=1; m<=mm; m++) YN[nn][m]=XN[n][m], YD[nn][m]=XD[n][m];
                for (Int_t m=1; m<=mm; m++) SumYN[m] += YN[nn][m], SumYD[m] += YD[nn][m];
        }

        for (Int_t m=1; m<=mm; m++) Ave[l][m] = SumYN[m]/SumYD[m];

        Int_t p = 2, zz = 2;

        Cum[l][1] = Ave[l][1];

        for (Int_t i=1; i<mm; i++){
                for (Int_t j=1; j<zz; j++){
                sACum[l][zz] += TMath::Binomial(zz,j)*TMath::Binomial(zz-1,j)*Ave[l][j]*Cum[l][zz-j];
                }
                Cum[l][p] = Ave[l][p] - sACum[l][zz]; // Q-cumulants given by Eq. (17)
                p += 1;
                zz = p;
        }

        for (Int_t u=1; u<=mm; u++){
                v2[l][u] = pow(1/aa[u]*Cum[l][u],1/(u*2.)); // v₂{2k} values given by Eq. (18)
        }

}

for (Int_t u=1; u<=mm; u++){
        Sumv2[u] = 0.0;
        for (Int_t h=1; h<=mm; h++){
                Sumprodv2[u][h] = 0.0;
        }
}

Int_t B = 0;
```

```
for (Int_t l=0; l<=Bootstrapps; l++){
        Int_t l2 = 0;
        for (Int_t u=1; u<=mm; u++){ // selecting only bootstrapps which give real values for all
v_2{2k}
                if (pow(-1,u)*Cum[l][u] < 0) l2 += 1;
        }

        if (l2==mm){
                B += 1;
                for (Int_t u=1; u<=mm; u++){
                        Sumv2[u] += v2[l][u];
                        for (Int_t h=1; h<=mm; h++){
                                Sumprodv2[u][h] += v2[l][u]*v2[l][h];
                        }
                }
        }
}

for (Int_t u=1; u<=mm; u++){
        for (Int_t h=1; h<=mm; h++){
                Cov[u][h] = (Sumprodv2[u][h] - Sumv2[u]*Sumv2[h]/B)/(B-1);
        } // covariance matrix of v_2{2k} for all k (1 < k < mm)
}

cout << fixed << setprecision(20) << endl;

for (Int_t u=1; u<=mm; u++){
        cout << " v_2{" << 2*u << "} = " << v2[0][u] << " +/- " << sqrt(Cov[u][u]) << endl;
}

cout << endl;

for (Int_t u=1; u<mm; u++){
        cout << fixed << setprecision(0) << " v_2{" << 2*u << "}/v_2{" << 2*mm << "} -1 = " << fixed
<< setprecision(20) << v2[0][u]/v2[0][mm] -1 << endl;
        cout << "                  +/- " << v2[0][u]/v2[0][mm]*sqrt(Cov[u][u]/pow(v2[0][u],2)-
2*Cov[u][mm]/(v2[0][u]*v2[0][mm])+Cov[mm][mm]/pow(v2[0][mm],2)) << endl; // for large order of
cumulants (mm), one eventually reaches the limit of digits by which the ROOT operates. This generates
false covariance matrix elements which in return gives the wrong uncertainties of the v_2{2k}/v_2{2mm}
ratios.
}
cout << endl;
}
```